\documentclass[twocolumn,showpacs,preprintnumbers,amsmath,amssymb]{revtex4}
\usepackage{graphicx}% Include figure files
\usepackage{dcolumn}% Align table columns on decimal point
\usepackage{bm}% bold math
\begin{document}
\title
{ Finite temperature effects in brane cosmology}%
\author{Rizwan Ul Haq Ansari}
\email{ph03ph14@uohyd.ernet.in}
\author{P K Suresh}
\email{pkssp@uohyd.ernet.in}
 \affiliation{ School of
Physics,University of Hyderabad.Hyderabad 500 046.India.}
\begin{abstract}
 Finite temperature effects in brane world cosmology are
 studied by  considering the interaction between scalar field and bulk
gravity. One-loop correction to   zero-temperature potential is
computed by taking into account,  interaction of scalar field and
bulk gravity. Phase transitions and high temperature symmetry
restoration  are examined. Critical temperature of  phase
transitions depends on the interaction constant of the scalar
field and bulk gravity, and these constant is an order parameter.
Present study can account  for second order phase transition in
early universe, in  brane world cosmological scenario.
\end{abstract}
\pacs {04.50.+h,98.80.k,98.80.Cq}
 \maketitle
\section{Introduction}
The  idea  of extra dimensions was proposed in the early twentieth
century  by Nordstorm and few years later by Kaluza and Klein
\cite{1}. It  has  reemerged  over  the  years  in  theories
combining  the principle  of  quantum  mechanics  and  relativity.
In particular theories   based  on  super  symmetry, especially
super  string theories,  are naturally expressed in more than four
dimensions\cite{2}. Recent  developments  in  string theory and
its extension M-theory have  suggested  another  approach  to
compactify  extra  spatial dimensions.  According  to  this, the
standard model particles are confined  on  a  hyper-surface,
called brane, embedded in a higher dimensional  space,  called
bulk. Only  gravity and other exotic matter such as dilaton can
propagate in the bulk. Our universe may be  such  a  brane like
object. This idea was originally motivated
phenomenological\cite{3,4,5,6,7} and later revival in string
theory by Horava and  Witten\cite{8}. Heterotic string theory
demand eleven dimensional super-gravity to  describe low energy.
Witten suggested that six of  eleven dimensions  can  be
consistently compactified on  a Calabi-Yau manifold.  Thus,  in
that limit space-time looks five dimensional with four dimensional
boundary brane\cite{9}. This provides the underlying  picture for
many brane world models proposed so far.

To solve the hierarchy problem, in the brane world scenario, it is
assumed  that  the  large extra dimensions may be compact but not
small\cite{10,11}. Recently, the idea   that   our  universe  is
embedded  in higher dimensional space-time has received much
attention \cite{11,12,13}  in  cosmology.  It  was proposed that
our observable universe          is          a three dimensional
 surface  (brane)  embedded  in a higher dimensional space (bulk).
Fundamental  structure  of  brane is well understood by two models
Dvali,   Gabbadadze   and   Porrati   model  (DGP)\cite{13} and
Randall-Sundrum  model(R-S  I and II)\cite{11,12}. In RS models,
one non-compact  extra  dimension is warped by a negative
cosmological constant,  hence  the  geometry  is warped. In DGP
model the extra dimension  is  flat  and  the  effect  of bulk
gravity on brane is considered  due  to interaction between brane
matter and graviton.

 The  introduction  of  branes into cosmology opened another novel
 approach to  understand the universe and  its evolution.
 This new perception of brane world opened new directions in cosmology,
 but  at the same time imposed some new problems. The cosmological
 evolution  of  the  universe  can  take  place  on brane, but
 for  the  whole  theory  to make sense, the brane should embedded
 in  a  consistent  way  to  higher  dimensional  space time, the bulk.
 The  only  physical field in the bulk is the gravitational field,
 and  there are no matter fields.

Brane world cosmology have been examined in problems like,
cosmological phase transitions \cite{14}, inflationary scenario
\cite{15}, baryogenesis \cite{16} stochastic background of
gravitational waves \cite{17}, singularity,flatness and entropy
problems \cite{18},
 induced gravity effects on branes  \cite{19}, quantum effects of
 bulk scalar field at non-zero temperature \cite{20},
in early universe.

In brane world scenario localized matter fields on the brane  can
couple to bulk gravity, which generate a localized  term for the
coupling \cite{13,21}.
 Therefore, if the brane world scenario is  correct for
cosmology, the effect of bulk on the evolution of  matter field,
is  also be taken into account.
In the present work, we examine phase transitions and symmetry
restoration occurred in early universe due to massive
scalar field which couple  with bulk
gravity, in the brane world cosmological scenario.

The idea of spontaneous symmetry breaking and its restoration is
useful to understand  the evolution of the early universe. In this
approach, it is believed that at high temperature, symmetries that
are spontaneously broken today were restored during the evolution of the
universe there were phase transition, perhaps many, associated
with the spontaneous breakdown of symmetries. In general, a
symmetry breaking phase transition can be first or second order.
If the phase transition is first order, the universe may be
dominated by the vacuum energy and undergo a period of inflation.
In this case, the transition proceeds by the nucleation of bubbles
of the true vacuum. If the phase transition is higher order the
thermal fluctuations may drive the transition.

 The    paper    is
 organized  as   follows. Section  I  gives  a  brief  introduction     for
 the  present work. Section II deals with the effects of bulk
 and     finite     temperature     on    the    scalar    field,
is considered and hence examined
the  phase transition and symmetry restoration in early universe.
Conclusions    and    discussions    are   presented   in   section
III.
\section{Phase Transition and Symmetry Restoration  }

 The
phase transitions studies due to effective potential  scalar field,
in the context of conventional four dimensional space time paradigm,
usually consider self interaction of the field. But
the scalar field interacts with other fields there can be
additional one-loop correction to the potential. Since bulk can
couple to matter field on the brane it would be useful to study its effects in  early
universe. In  present study, we consider a term  arise from
 the interaction of scalar field and bulk gravity which is on the
brane.

 The Lagrangian
which describe the coupling of massive scalar field with the bulk gravity can be written as
\begin{eqnarray}
L&=&{1\over2}\partial_A\phi\partial^A\phi-V(\phi,y),
\end{eqnarray}
where $A$=0,1,2,3,4 ($t=0,x^i=1,2,3~,y=4$)and $y$ is the
extra-dimensional coordinate. The above Lagrangian is assumed
invariant under $Z_2$ symmetry.

Since the scalar field coupled with  bulk gravity the
potential, $V(\phi,y)$, takes the following form
\begin{eqnarray}
V(\phi,y)&=&-{1\over2}m^2\phi^2+{1\over4}\lambda\phi^4+{1\over2}{\cal{D}}\delta(y)\phi^2,
\end{eqnarray}
where $\lambda$ is the coupling constant due to self interaction
of field $\phi$. The third term in Eq.(2) is to account for the
interaction between the scalar field and bulk gravity, which is
assumed on the brane. Where
 ${\cal{D}}$ is the interaction constant  and  $\delta(y)$ due to
the fact that the coupling  confined on the brane.
Since   effects of the potential are seen on the brane, Eqn (2) to
be considered at $y=0$ for the further study.

Consider the potential(2), therefore we find,
\begin{eqnarray}
V(\phi_\pm)&=&-{1\over4\lambda}[m^2-{\cal{D}}]^2,
\end{eqnarray}
 and
 \begin{eqnarray}
V''(\phi_\pm)&=&2m^2-2{\cal{D}}.
\end{eqnarray}
 Where prime denotes differentiation with respect to $\phi$ and $\phi_\pm=\pm
\left(m^2-{\cal{D}}\over \lambda\right)^{1/2}$ are the two
equivalent minima. The discrete symmetry of the lagrangian is
broken by choosing either of the vacuum state \cite{20}and the mass of
physical boson is determined by curvature of the potential about
the true ground state $M^2=V''(\phi_\pm)=2m^2-2{\cal{D}}=2\lambda \phi_\pm^2$.

Our, next aim is to study the finite temperature effects and
symmetry restoration due to the potential Eq.(2).
In order to study phase transitions in  the  early universe, the
corresponding effective potential of the scalar field to  be
computed at finite temperature. Thus finite temperature effective
potential on field theory can be used to study the phase
transition in the early universe, which in turn shows  symmetry
breaking present in a model under consideration.

The potential given by Eq.(2) is the temperature independent
zero-loop effective potential. The finite temperature effective
potential($V_T$) is the free energy density associated with field.
The one-loop approximation to the to the finite temperature
 potential ($V_{1}^T$)can be computed as the method of Dolan and
Jackiw \cite{21} and  is given by
\begin{eqnarray}
V_{1}^{T}&=&\frac{ T}{2}\sum_{n}\int\frac{d^3k}{(2\pi)^3}
\ln( k^2 - M^2)\\
&=&\frac{T}{2}\sum_{n}\int\frac{d^3k}{(2\pi)^3} \ln( -4
 \pi^2 n^2 T^2- \epsilon^2_{M}),
\end{eqnarray}
where
\begin{eqnarray}
\epsilon^2_{M}&=&k^2+ M^2,\\ \nonumber
 M^2(\phi)&=&-m^2+ {\cal{D} }+ 3\lambda\phi^2.
\end{eqnarray}
Following Dolan and Jackiw \cite{21}, the sum on $n$ diverges can
be computed. Thus we find
\begin{eqnarray}
V_1 ^T =V_{1}^{0}+ \bar{V}_{1} ^T.
\end{eqnarray}
Where
\begin{eqnarray}
V_{1}^{0}=-\frac{1}{2}m^2 \phi^2 +\frac{\lambda}{4}\phi^4+
{{\cal{D}}\over2}\phi^2+
\frac{M^4}{64\pi^2}\ln(\frac{M^2}{\mu^2}),
\end{eqnarray}
is the one-loop temperature independent effective potential and
 $\mu$ is an arbitrary mass scale and can be
related to renormalization constants. In Eq(8) $\bar{V}_{1}^T$ can
be written as  in high temperature limits \cite{21}as follows
\begin{eqnarray}
\bar{V}_{1} ^T=-\frac{\pi^2}{90}T^4+ \frac{M^2}{24}
T^2-\frac{M^3}{12 \pi} T-\frac{M^4} {64 \pi^2}\ln \frac{M^2}{T^2}.
\end{eqnarray}
 The symmetry restoration can be achieved if
the temperature is raised above a certain temperature  known as
the critical temperature. The critical temperature in the present
case is obtained as
\begin{eqnarray}
T_c=\sqrt{ 4 \frac{m^2-{\cal{D}}}{\lambda}}.
\end{eqnarray}
The critical temperature  vary depending upon the value of
${\cal{D}}$ and the phase transition can occur according to it.
Critical temperatures for different values of ${\cal{D}}$  are
presented in  Table.1.

The behavior of the potential $V_T$ is shown in Fig1.a,b,c,d,e for
 fixed value of m,$\lambda$,${\cal{D}}$
and different temperatures. Behavior of the effective potential for different
${\cal{D}}$ is shown in Fig2 for a given temperature.
\begin{figure}
\input epsf
\centerline{ \epsfxsize=2.25 in \epsfbox{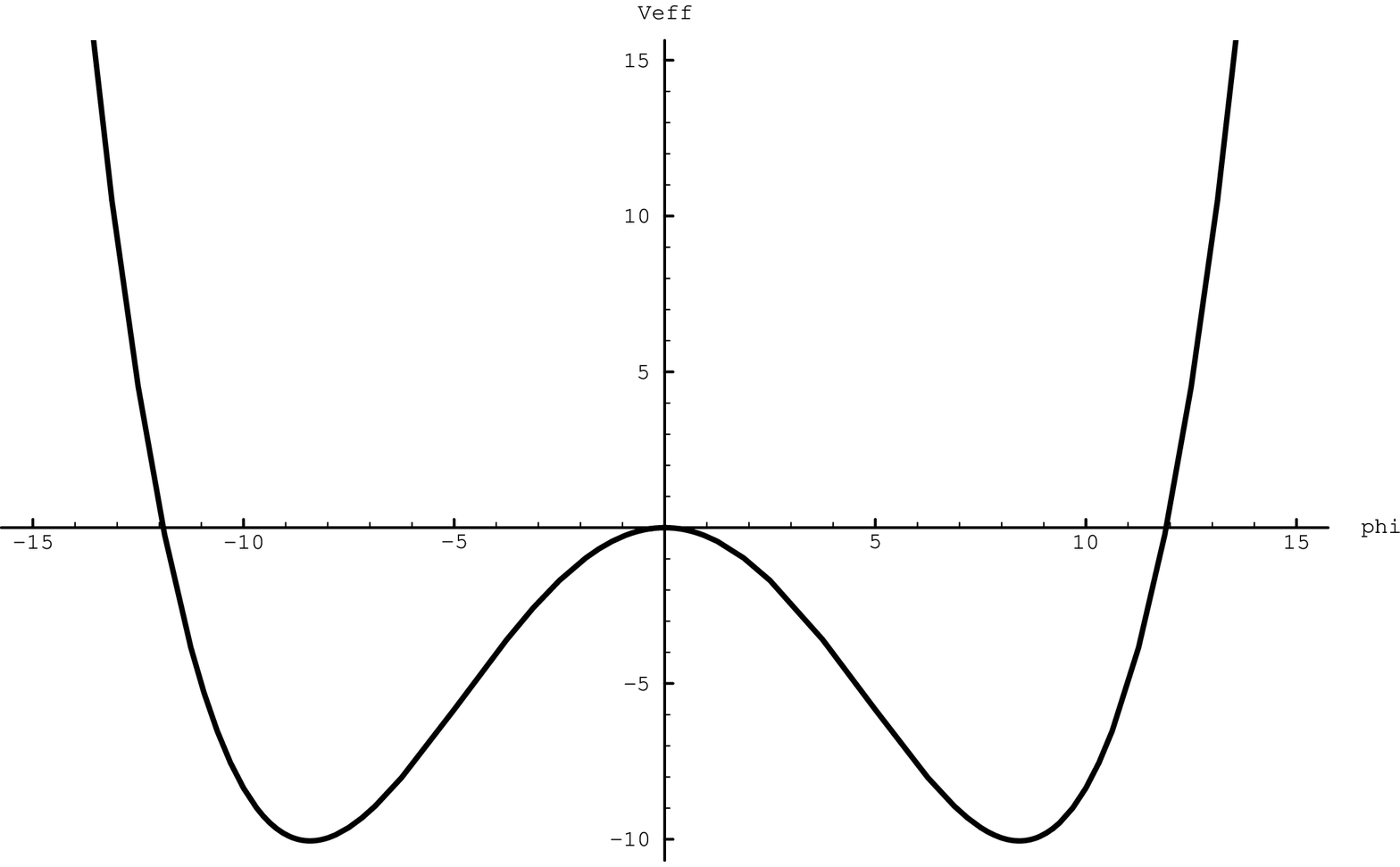}}
\vspace{-1.75cm}
 FIG :1a
 \end{figure}
 \begin{figure}
\input epsf
\centerline{ \epsfxsize=2.25 in \epsfbox{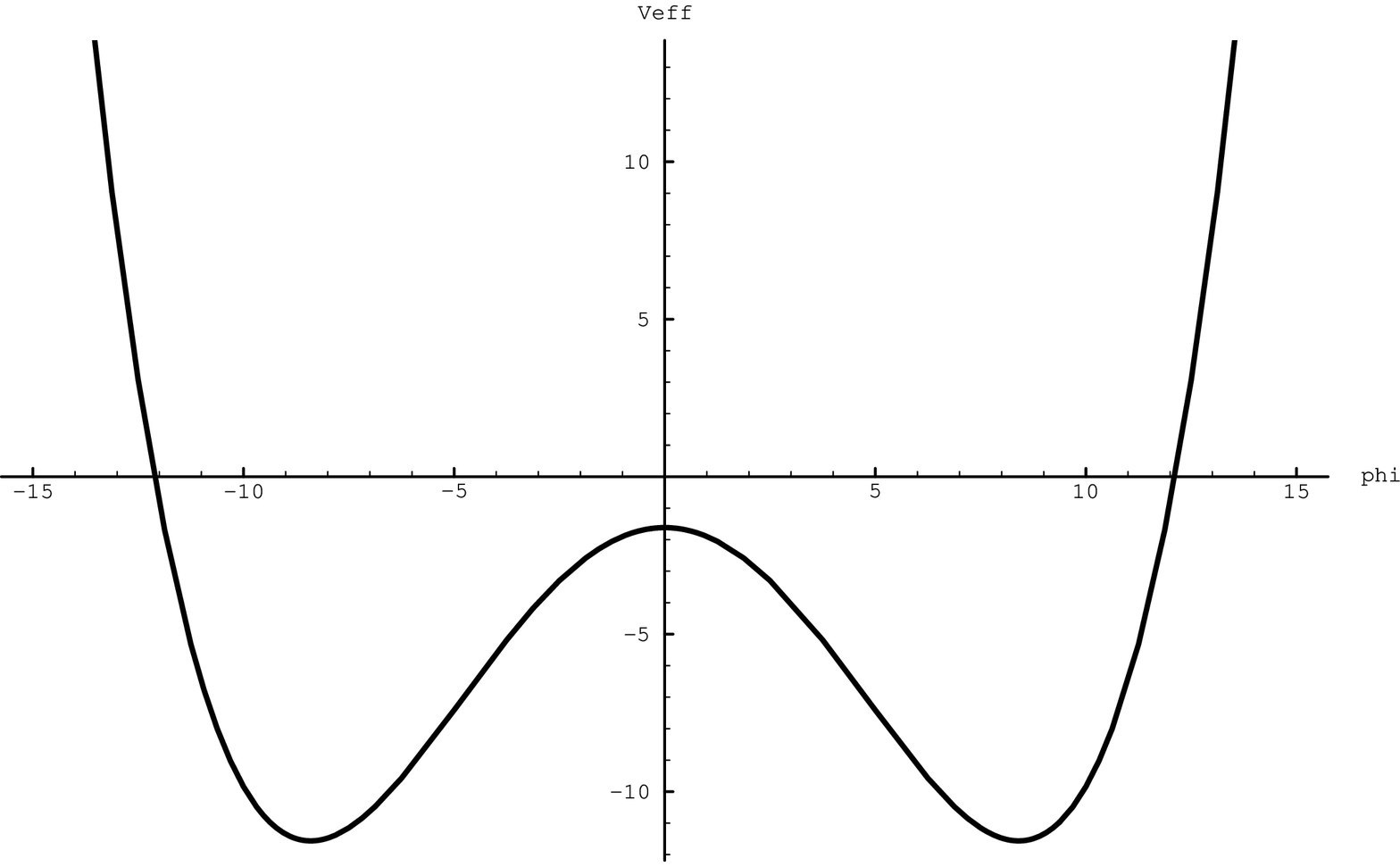}}
\vspace{-1.75cm}
 FIG :1b
 \end{figure}
\begin{figure}
\input epsf
\centerline{ \epsfxsize=2.25 in \epsfbox{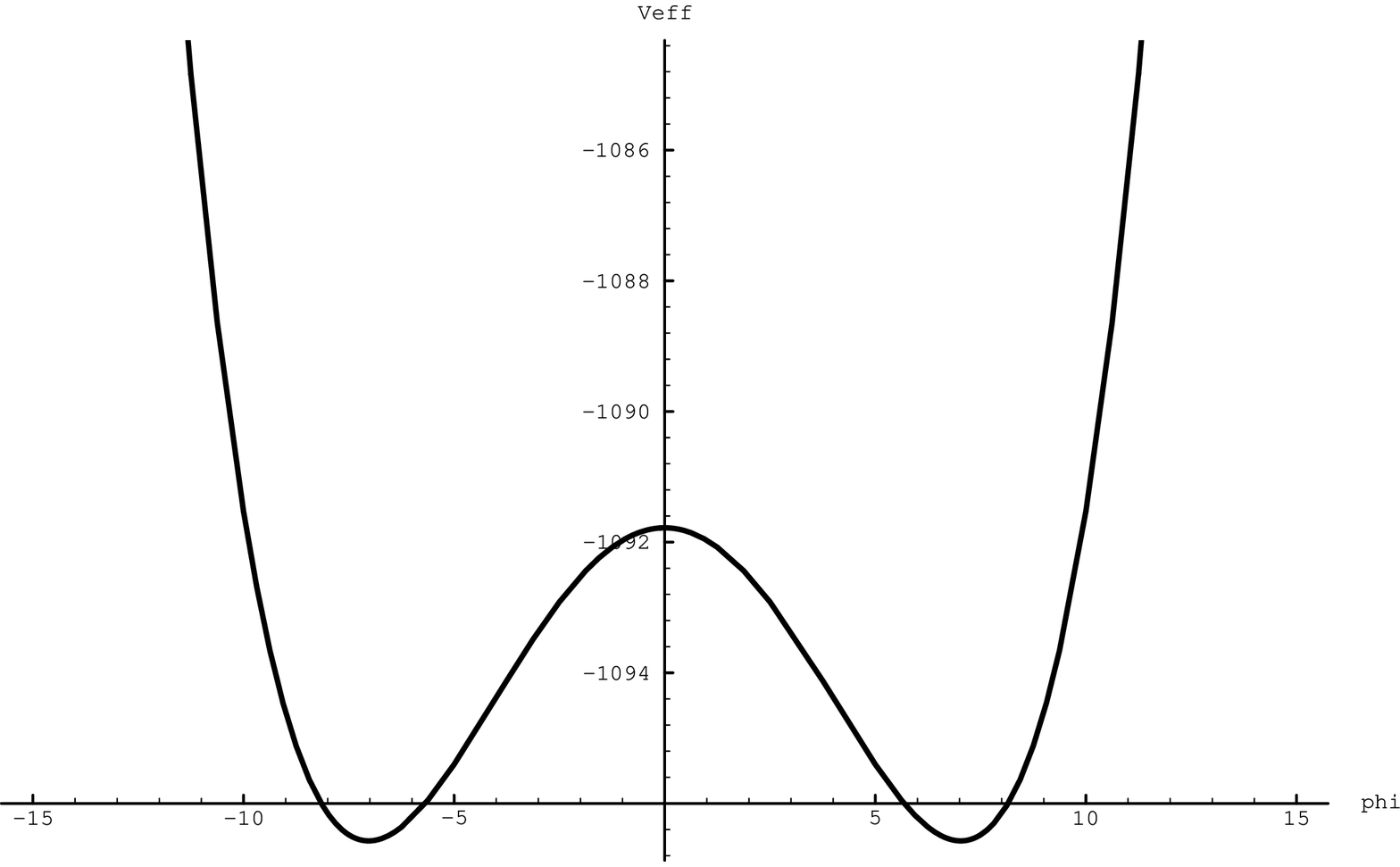}}
\vspace{-1.75cm}
 FIG :1c
 \end{figure}
 \begin{figure}
\input epsf
\centerline{ \epsfxsize=2.25 in \epsfbox{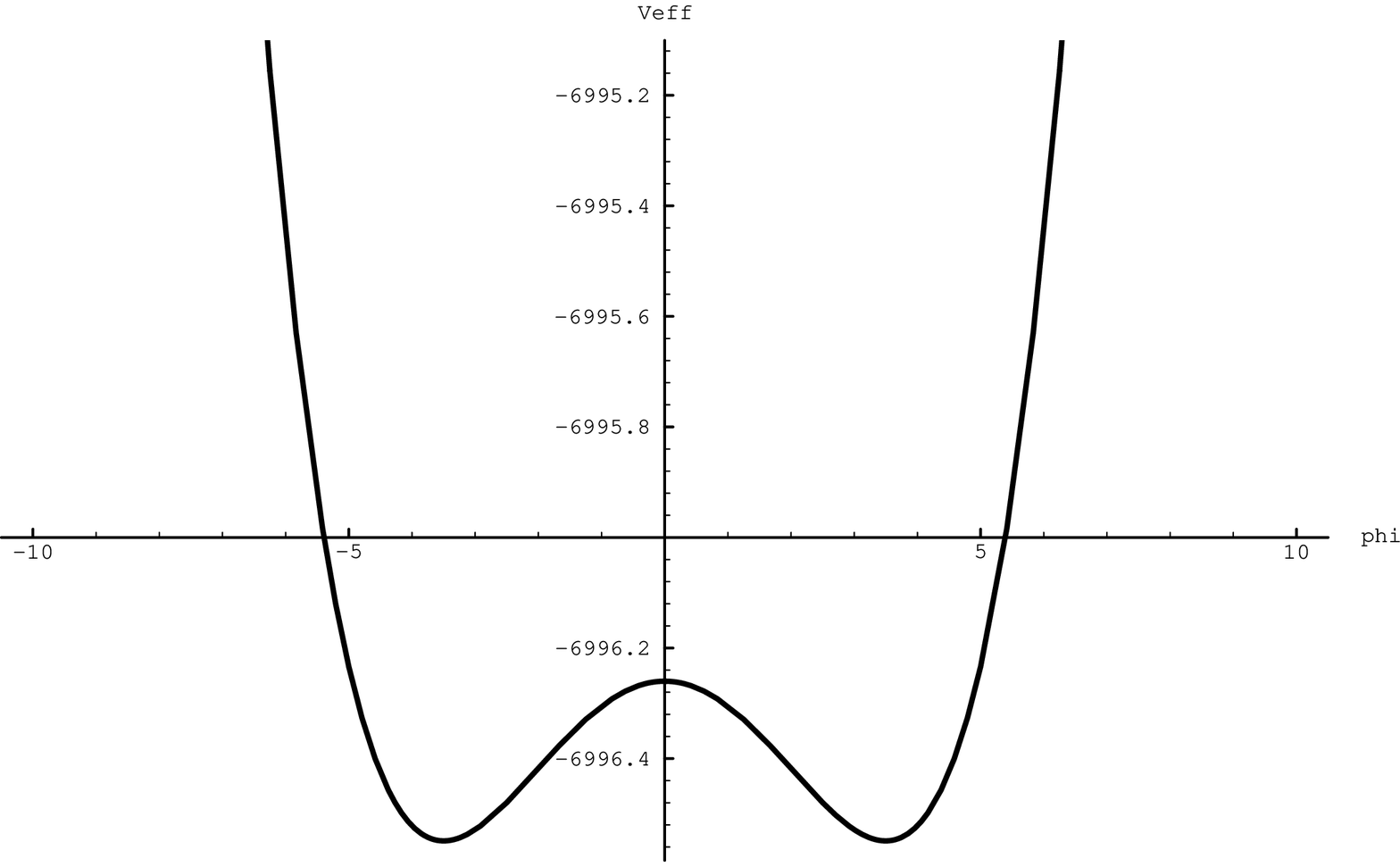}}
\vspace{-1.75cm}
 FIG :1d
 \end{figure}
 \begin{figure}
\input epsf
\centerline{ \epsfxsize=2.25 in \epsfbox{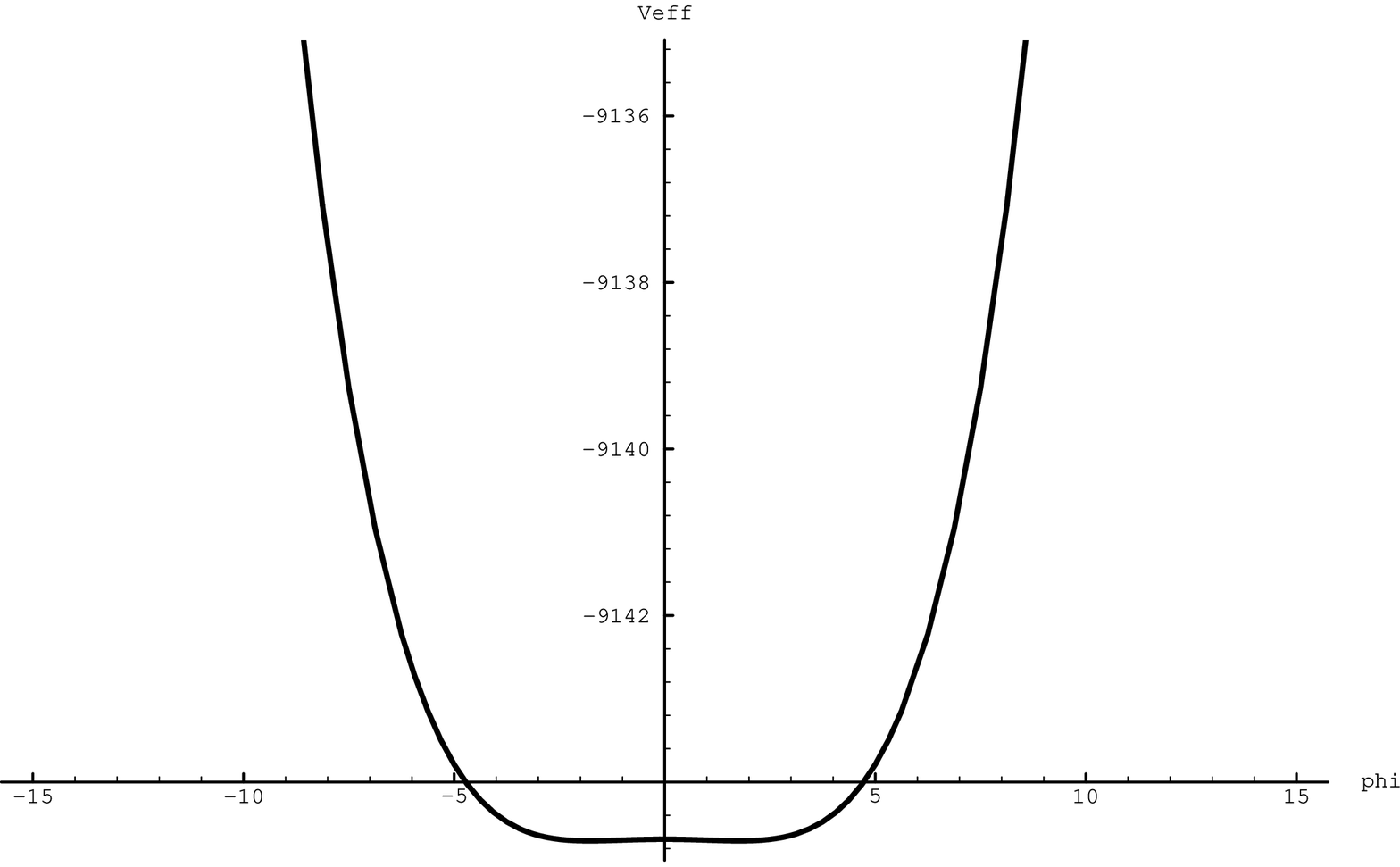}}
\vspace{-1.75cm}
 FIG :1e
 \end{figure}
 \begin{table}
 \caption{}
 Critical temperature $T_{c}$ for corresponding
${\cal{D}}$, for fixed $m=0.9371$ and $\lambda =0.008$\\
 \begin{ruledtabular}
\begin{tabular}{|c|c|c|c|}
  \hline
  ${\cal{D}}$ & $T_c$ & ${\cal{D}}$& $T_c$ \\
  \hline
  0.7 & 9.43 & 0.005 & 20.89 \\
  \hline
  0.5 & 13.75 & 0 & 20.95 \\
  \hline
  0.3 & 17.00 & -0.1 & 22.11 \\
  \hline
  0.1 & 19.72 & -0.3 & 24.27 \\
  \hline
  0.05 & 20.34 &-0.5 & 26.25 \\
  \hline
\end{tabular}
 \end{ruledtabular}
 \end{table}
\begin{figure}
\input epsf
\centerline{ \epsfxsize=2.25 in \epsfbox{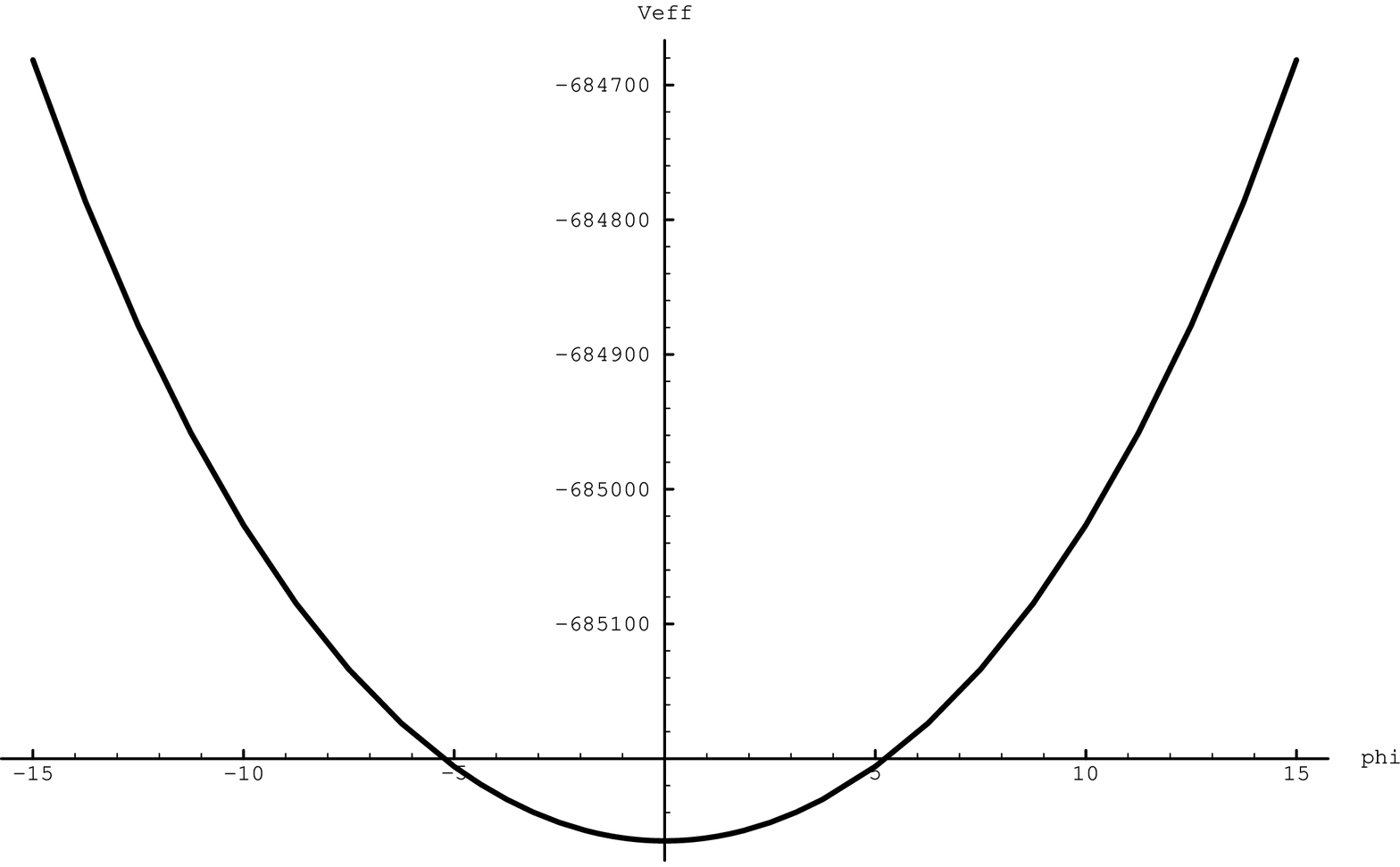}} \vspace{-2cm}
 FIG :1f
 \caption{ a,b,c,d,e show
the behavior of the effective potential $V_T$  for
 $ m=0.9371$,$\lambda=0.008$,and ${\cal{D}}=0.37$
$T= 0,2,10,15.9,17,50$ respectively }
 \end{figure}
 \begin{figure}
\input epsf
\centerline{ \epsfxsize=2.25 in \epsfbox{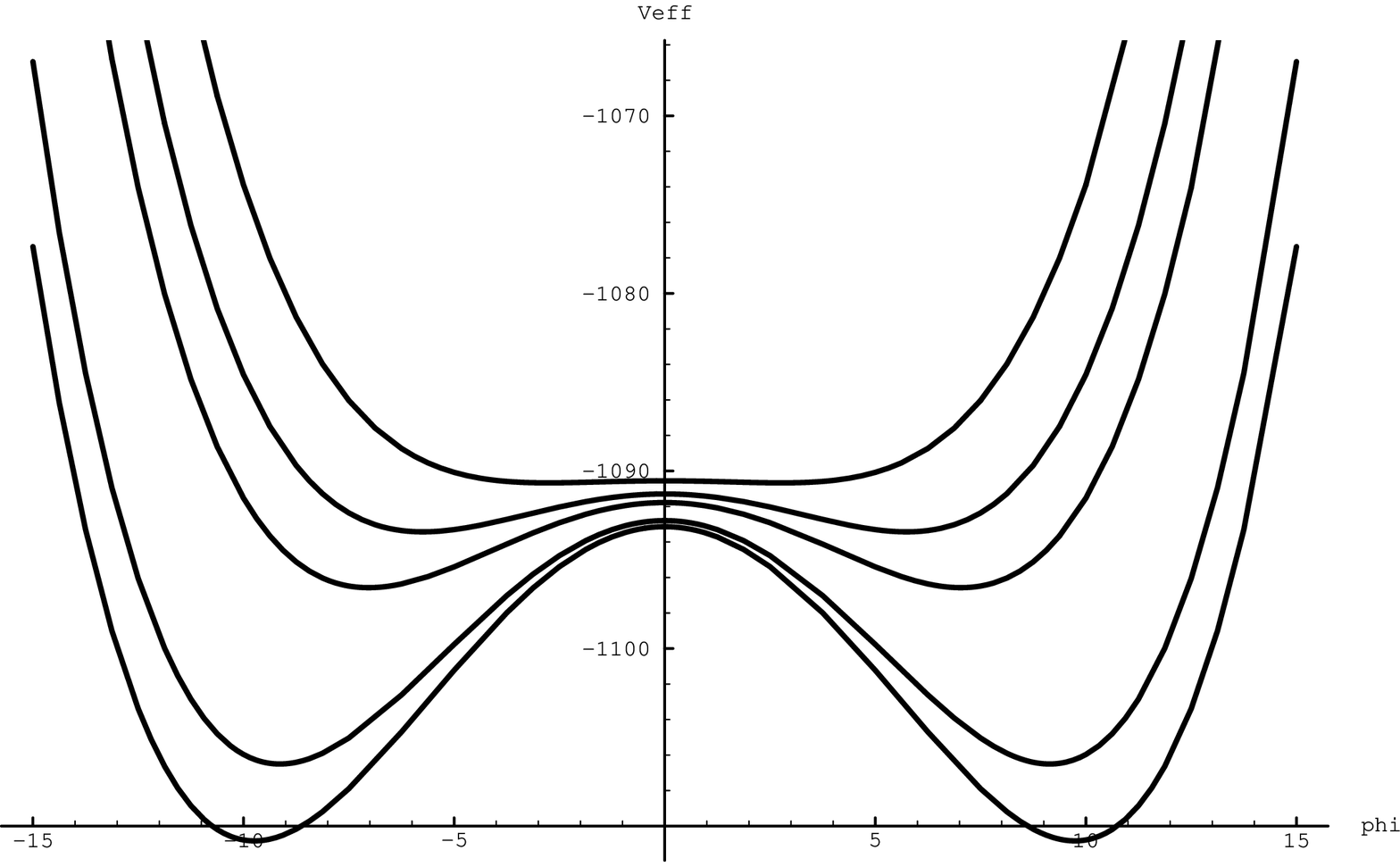}} \vspace{-2cm}
 \caption{Shows the behavior of the effective potential for
${\cal{D}}= 0.7,0.5,0.37,0.1,0.01$, $m=0.9371,\lambda =0.008$  and $T=10$.}
\end{figure}
\section{Conclusions }
In this paper we studied finite temperature effects in brane world
cosmology by  considering the interaction between scalar field and
bulk gravity. One-loop correction to the  zero-temperature
potential is considered by taking into account of the interaction
of scalar field and bulk gravity. Hence the phase transition and
high temperature symmetry restoration in brane world scenario is
examined. The behavior of finite temperature effective potential
is analyzed. Fig1.a,b,c,d,e show  behavior of the potential $V_T$
for fixed value of m,$\lambda$,${\cal{D}}$ and different
temperatures.
At $T=0$ the symmetry of finite temperature effective potential
is broken (FIG1a). The behaviour of the potential
at the critical temperature,$T_C=15.9$, is shown in FIG1.c.
When $T<<T_c$ and $T>T_c$, the corresponding behaviour of the
potential are shown in FIG1.b and FIG1.e respectively.
 When $T>>T_c$, the  symmetry of the potential became resorted (FIG1.f).
 From the plots it can be see that   at the critical temperature
 the transition occurs smoothly, which is the nature of
 second order phase transition. Hence from
these plots, it is evident that the nature of phase transition
 in the present study is second order.
From the Table.1  it can be concluded that the critical
temperature, $T_c$, dependent on  the interaction
constant,${\cal{D}}$, due to the scalar field and bulk gravity.
Plots for the finite temperature effective potential with the scalar field for
various values of ${\cal{D}}$ are shown in FIG2 for fixed values of
$m,\lambda $and $T$,which has the signature of second order phase transition.
These plots show that the symmetry restoration is also dependent on ${\cal{D}}$, hence it
can be an order parameter for the phase transition. The present
study can account  for  second order phase transition in the early
universe, which is due to the interaction of scalar field and
bulk,in the brane world scenario. This changes in phase transition
due ${\cal{D}}$ may have important consequences on inflation. The
second order phase transition is very useful to understand the
dynamics of very early universe. When $ {\cal{D}} $=0, the
present results reduce to the self interacting scalar field, which is
minimally coupled with gravity, in the conventional four
dimensional space time paradigm.
We hope that the idea used in the present work may be useful to
understand inflationary scenario and related issues in early
universe,in brane world scenario.

\end{document}